\documentclass[Phys]{SciPost}

\hypersetup{
    colorlinks,
    linkcolor={red!50!black},
    citecolor={blue!50!black},
    urlcolor={blue!80!black}
}

\usepackage[bitstream-charter]{mathdesign}
\DeclareSymbolFont{usualmathcal}{OMS}{cmsy}{m}{n}
\DeclareSymbolFontAlphabet{\mathcal}{usualmathcal}

\begin{document}

\begin{center}{\Large \textbf{
Workshop on Tau Lepton Physics: $\mathbf{30^{th}}$ Anniversary\footnote[4]{Dedicated to the memory of Simon Eidelman and Olga Igonkina}
\\
}}\end{center}

\begin{center}
Antonio Pich 
\end{center}

\begin{center}
Departament de Física Teòrica, IFIC, Universitat de València – CSIC\\
Parque Científico, Catedrático José Beltrán 2, E-46980 Paterna, Spain
\\
Antonio.Pich@ific.uv.es
\end{center}

\begin{center}
\today
\end{center}


\definecolor{palegray}{gray}{0.95}
\begin{center}
\colorbox{palegray}{
  \begin{minipage}{0.95\textwidth}
    \begin{center}
    {\it  16th International Workshop on Tau Lepton Physics (TAU2021),}\\
    {\it September 27 – October 1, 2021} \\
    \doi{10.21468/SciPostPhysProc.?}\\
    \end{center}
  \end{minipage}
}
\end{center}

\section*{Abstract}
{\bf
The first Workshop on Tau Lepton Physics took place at Orsay in 1990.
The evolution of the field and some physics highlights are briefly described, following the presentations discussed at the fifteen $\boldsymbol{\tau}$ workshops
that have been held  since then. 
}



\begin{figure}[h]
\centering
\includegraphics[width=0.45\textwidth]{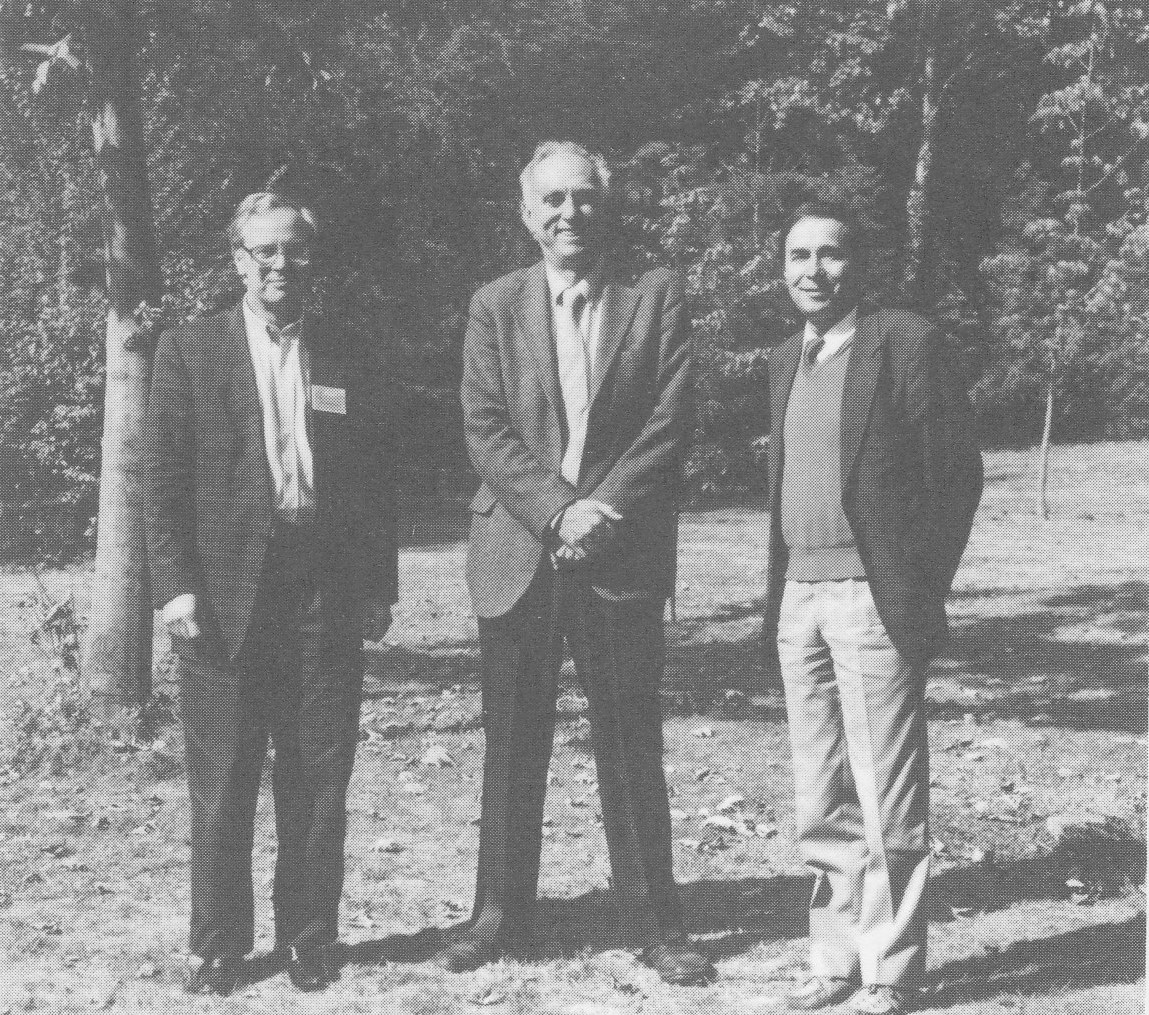}
\caption{Bernard Jean-Marie (left), Martin Perl (center) and Michel Davier (right) at the first workshop on Tau Lepton Physics (Orsay, 1990) \cite{Davier:1991jq}.}
\label{fig:mass}
\end{figure}


In 2020 we were supposed to celebrate the $30^{\mathrm{th}}$ anniversary of the
Workshop on Tau Lepton Physics, a very successful series of scientific meetings \cite{Davier:1991jq,Gan:1992fum,Rolandi:1995yx,Smith:1997iq,PichZardoya:1999yv,Sobie:2001it,Schalk:2002zs,Ohshima:2005iv,Cei:2007zza,Bondar:2009zz,Lafferty:2011zz,Hayasaka:2014xya,Stahl:2015oci,Yuan:2017tgx,Proceedings:2019ihn}
that was initiated in 1990, at Orsay, by Michel Davier and Bernard Jean-Marie \cite{Davier:1991jq}.
Owing to the covid pandemic, the $16^{\mathrm{th}}$ Tau Workshop and this celebration have finally taken place on-line, after a one-year delay.
Meanwhile, the $\tau$ community was shocked with the sad losses of our friends Simon Eidelman and Olga Igonkina, summary speaker and main organizer, respectively, of the TAU 2018 Workshop \cite{Proceedings:2019ihn}, who are deeply missed by all of us.

In the following sections, I describe the 1990 status and the posterior evolution of the field, through a selection of physics highlights.
Obviously, I cannot cover the large number of excellent contributions discussed in the fifteen workshops organised so far. I apologize in advance for the many omissions in this incomplete and very personal (subjective) overview. A long list of relevant references can be found in my 2014 review on $\tau$ physics \cite{Pich:2013lsa}.

\section{Orsay 1990: a Successful and Inspiring Meeting}

Since its discovery in 1975 \cite{Perl:1975bf} at the SPEAR $e^+e^-$ storage ring,
the properties of the  $\tau$ lepton were investigated by many experiments, finding quite reasonable agreement with the pioneering predictions of Yung-Su Tsai \cite{Tsai:1971vv}. The status before the Orsay meeting was nicely summarized in 1988 by Barry C. Barish and Ryszard Stroynowski in an extensive Physics Report  \cite{Barish:1987nj}, and the situation was far from being satisfactory. Tau physics was not yet a field on its own, since it had just developed as a ``by-product of general-purpose $e^+e^-$ detectors'' (B. Barish, in \cite{Davier:1991jq}). The $\tau$ data had relatively large uncertainties and
exhibited internal inconsistencies. In particular, many efforts were being devoted to understand two long-standing anomalies: 
\begin{enumerate}
\item 
The so-called missing one-prong problem, a $3.3\,\sigma$ deficit of the sum of exclusive $\tau$-decay branching ratios into channels with one single charged particle, $\mathrm{B}_1^{\mathrm{excl}} = (81.0\pm 1.5)\%$, with respect to the inclusive measurement 
$\mathrm{B}_1^{\mathrm{incl}} = (86.1\pm 0.3)\%$. 
\item 
A sizeable $2.7\,\sigma$ discrepancy between the 
$\tau$ lifetime, $\tau_\tau^{\mathrm{exp}} = (3.04\pm 0.06)\cdot 10^{-13}$~s, and its Standard Model (SM) prediction, 
$\tau_\tau^{\mathrm{SM}} = (2.81\pm 0.06)\cdot 10^{-13}$~s, obtained from the measured branching ratio of the decay $\tau\to\nu_\tau e\bar\nu_e$, $\mathrm{B}_e^{\mathrm{exp}} = (17.78\pm 0.32)\%$. 
\end{enumerate}
A much larger anomaly, the 1987 HRS claim \cite{Derrick:1987sp} of a huge 5\% $\tau^+$ branching ratio into the G-parity violating $\eta\pi^+\bar\nu_\tau$ final state, had just been dismissed on experimental grounds.

A workshop devoted to study the physics potential of a low-energy $\tau$-charm factory ($\tau$cF) was organised in 1989 at SLAC \cite{Proceedings:1989lya}, triggering a renovated interest in this type of physics. This was followed by several other $\tau$cF meetings at different sites (Spain, USA, China\ldots) that would later culminate with the BEPCII project in Beijing. However, with the available luminosity, charm and charmonium were the clear physics priorities of this $\tau$cF. 

The start of the LEP operation in August 1989 was the main motivation to organise a topical workshop fully devoted to the $\tau$ lepton. Although the initially planned scientific programme of LEP had not paid much attention to $\tau$ physics, it was soon realised that this $e^+e^-$ collider was an excellent environment to perform precise measurements of the $\tau$ properties.
An increasing interest in this third-generation particle was clearly manifested by the large number (117) of participants attending the Orsay meeting, which triggered many ideas, suggestions and lively discussions. First, very preliminary, analyses of the LEP data were already presented (H.-S. Chen, D.E. Klem, G.G.G. Massaro, S. Orteu, S. Snow, A. Stahl, P. Vaz, M. Winter, F.~Zomer), showing the high physics potential of the new collider. In the Orsay proceedings \cite{Davier:1991jq}, one can already find first studies on topics that would later become common ingredients of the $\tau$ research: isospin relation with $e^+e^-$ data (S. Eidelman and V. Ivanchenko), polarization analysers (A.~Rougé), resonance studies (J.H.~K\"uhn), etc. I was invited to 
discuss a possible determination of the strong coupling ($\Lambda_{\overline{\mathrm{MS}}}$) from the inclusive $\tau$ decay width, suggested by Stephan Narison and myself \cite{Narison:1988ni}. 
It was a quite bold and heterodox proposal at the time, and my talk was finally scheduled in the new-physics section.

\section{The Golden Age}

\begin{figure}[h]
\centering
\includegraphics[width=0.45\textwidth]{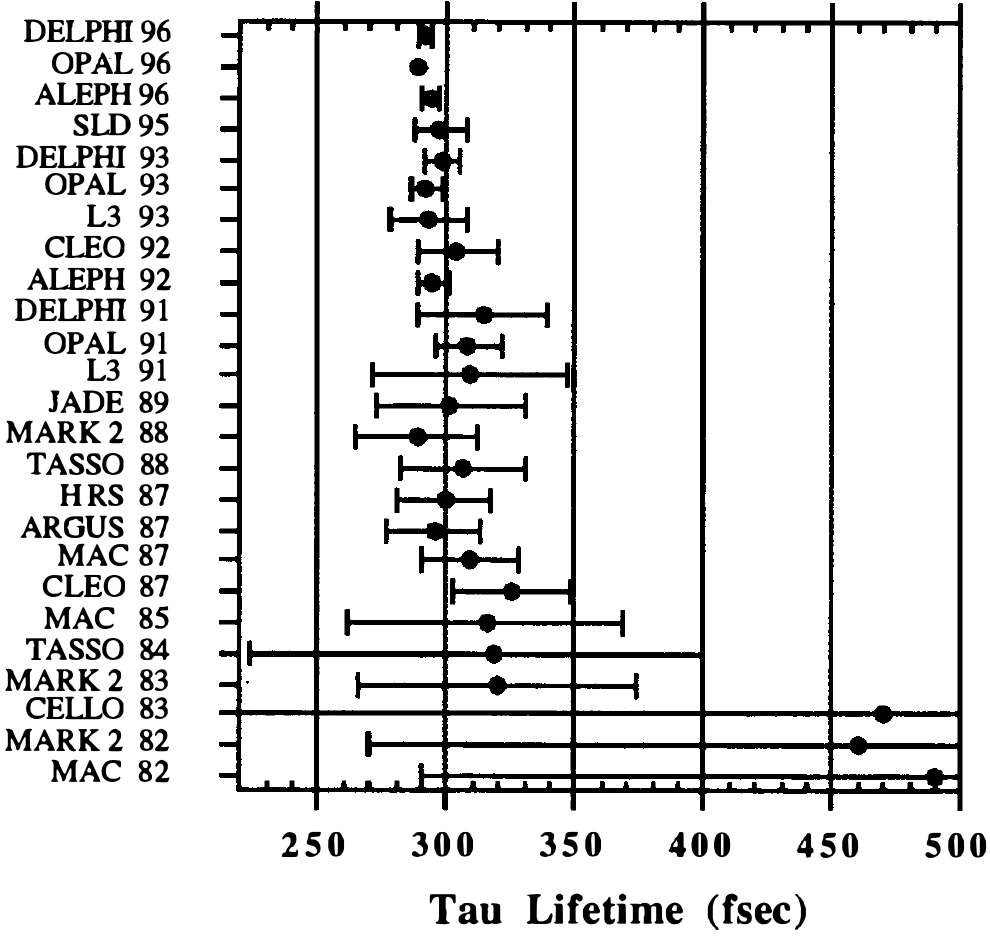}
\hskip 1cm
\includegraphics[width=0.45\textwidth]{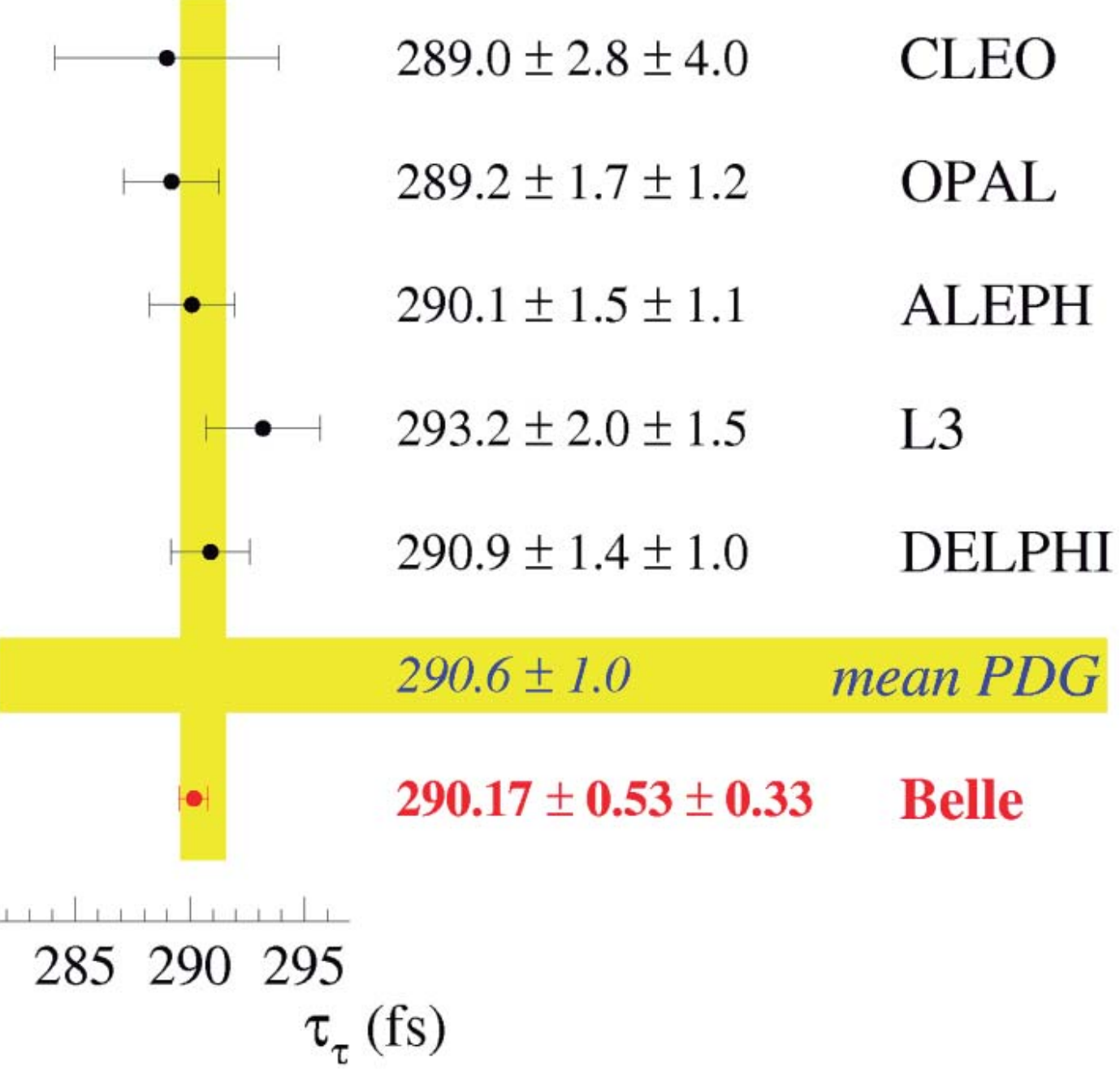}
\caption{$\tau_\tau$ status
in 1996 (K.G. Hayes in \cite{Smith:1997iq}) and 2014 (M.~Shapkin in \cite{Stahl:2015oci}).}
\label{fig:lifetime}
\end{figure}

\begin{figure}[h]
\centering
\includegraphics[width=0.55\textwidth]{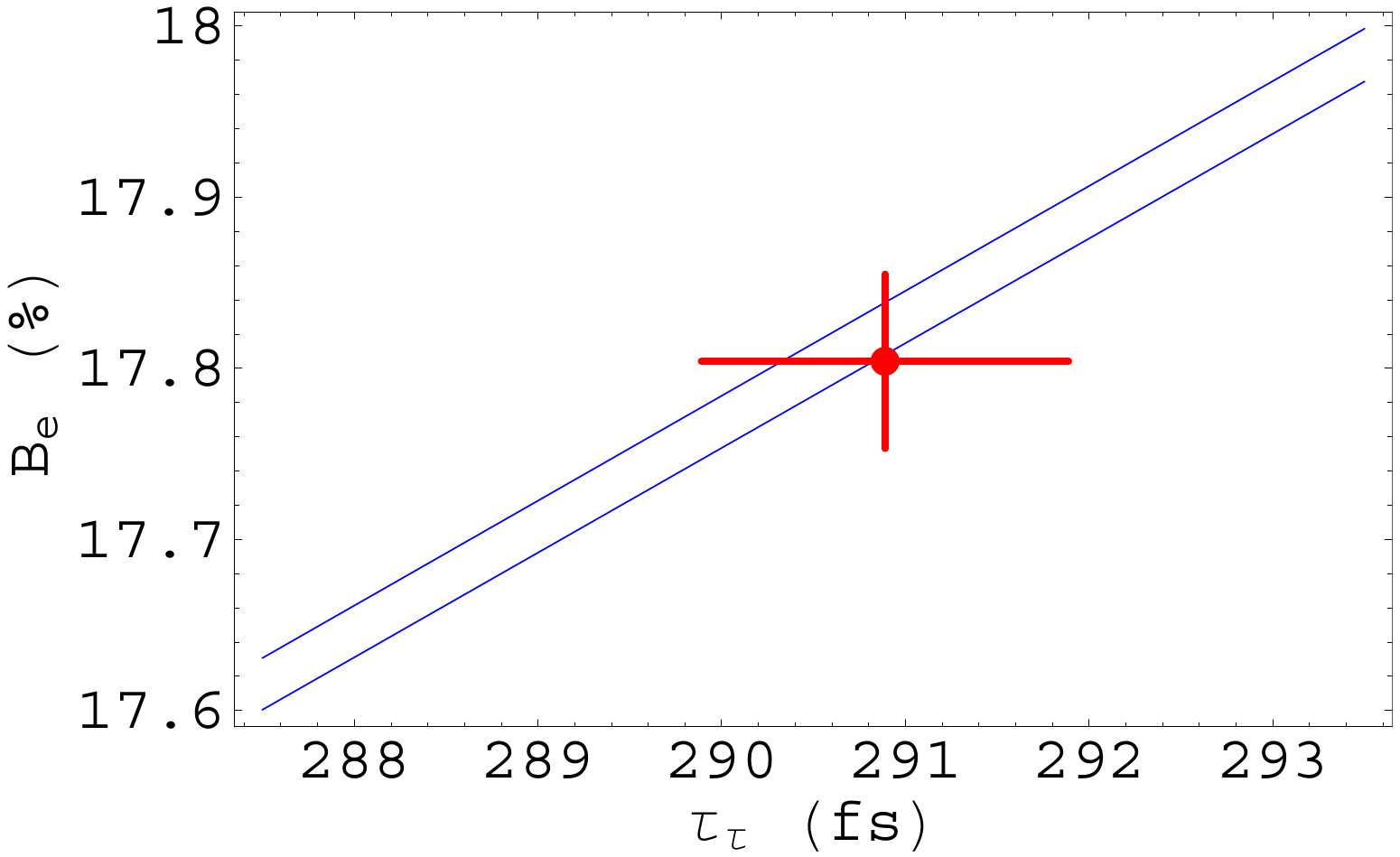}
\caption{2000 world averages of $\tau_\tau$ and $\mathrm{B}_e$. The blue lines show the SM prediction for the measured range of $m_\tau$ (A. Pich in \cite{Sobie:2001it}).}
\label{fig:BeLifetime}
\end{figure}

The next workshops on Tau Lepton Physics (Columbus 1992 \cite{Gan:1992fum}, Montreux 1994 \cite{Rolandi:1995yx}, Estes Park 1996~\cite{Smith:1997iq}, Santander 1998 \cite{PichZardoya:1999yv}, Victoria 2000 \cite{Sobie:2001it}) witnessed a fast and drastic qualitative change on the status of $\tau$ physics, with lots of good data coming from CLEO, LEP and SLD, together with the last ARGUS analyses.
In 1992, the disturbing $\tau$ anomalies were already solved (M.~Davier, W.J.~Marciano in \cite{Gan:1992fum}). A tight (95\% CL) limit on unmeasured decay modes was set with the LEP data, $\mathrm{B}_{\mathrm{unseen}}< 0.11\%$  (M.~Davier in \cite{Gan:1992fum}), and BES released a very precise measurement of the $\tau$ mass (H.~Marsiske in \cite{Gan:1992fum}), slightly below the previous world average. It was followed by a precise ALEPH measurement of the $\tau$ lifetime, subsequently confirmed by the other LEP detectors and SLD, which shifted $\tau_\tau$ to smaller values (M. Davier in \cite{Rolandi:1995yx}).  Those measurements are compared with their 2014 values in Figs.~\ref{fig:mass} and \ref{fig:lifetime}.
The current PDG averages are not much different:
$m_\tau = (1776.86 \pm 0.12)~\mathrm{MeV}$ and
$\tau_\tau = (290.3\pm 0.5)~\mathrm{fs}$ \cite{ParticleDataGroup:2020ssz}. The combination of these two experimental inputs eliminated the previous discrepancy with the electronic branching ratio, as shown in Fig.~\ref{fig:BeLifetime} (A. Pich in \cite{Sobie:2001it}).

\begin{figure}[h]
\centering
\includegraphics[width=0.45\textwidth]{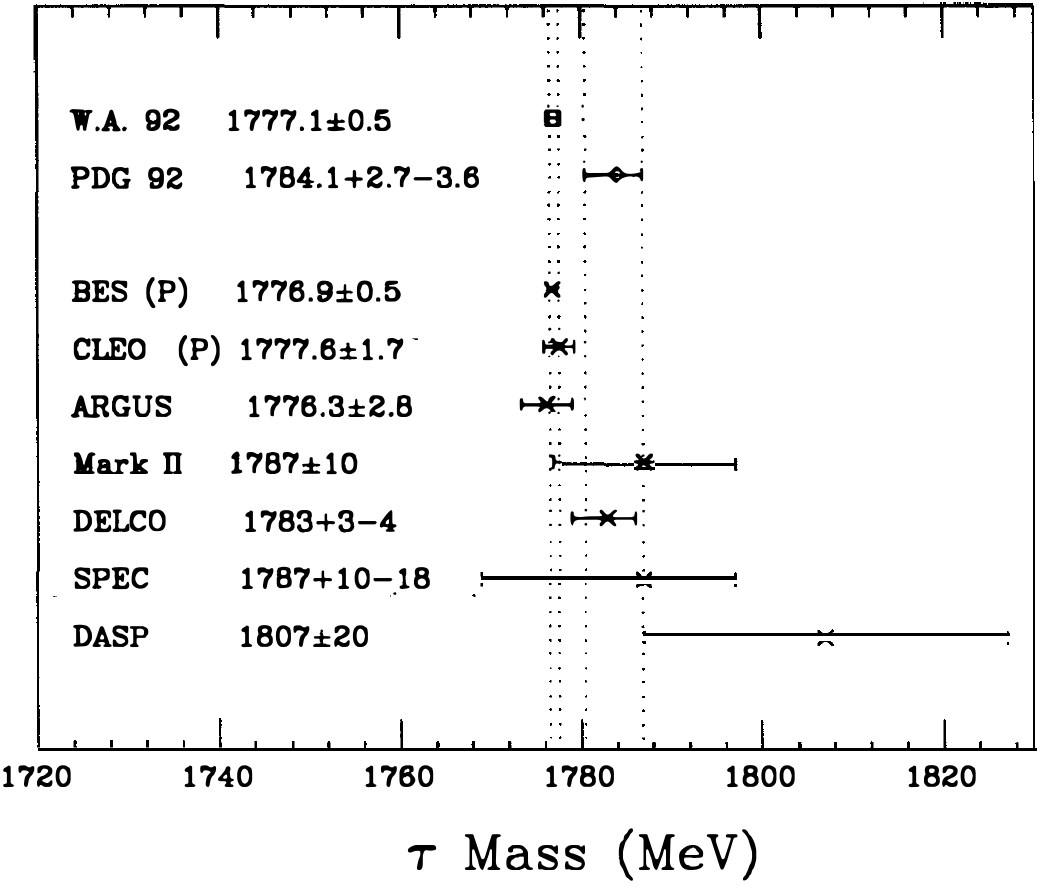}
\hskip .5cm
\includegraphics[width=0.5\textwidth]{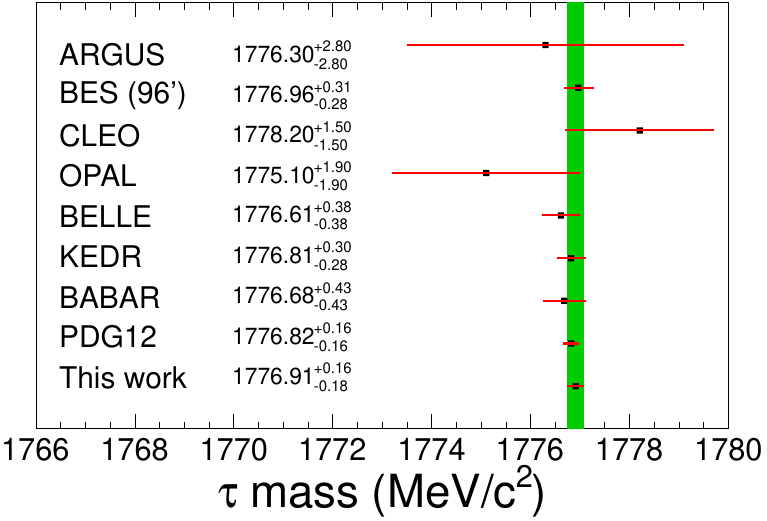}
\caption{$m_\tau$ status in 1992 (H. Marsiske in \cite{Gan:1992fum}) and 2014 (T. Luo in \cite{Stahl:2015oci}).}
\label{fig:mass}
\end{figure}

The excellent quality of the LEP data brought a new era of precision physics, which was complemented with many theory contributions, allowing us to perform accurate tests of the SM, in both the electroweak and QCD sectors. Around 2000, lepton universality was tested with a 0.2\% precision for charged currents (A. Pich in \cite{Sobie:2001it}), 
and the measured leptonic $Z$ couplings already indicated that low values of the Higgs mass were favoured (D.W. Reid in \cite{Sobie:2001it}).
Thanks to the $\tau$ polarization emerging from the decay $Z\to\tau^+\tau^-$, it was possible to analyse the Lorentz structure of the leptonic $\tau$ decays and put relevant bounds on hypothetical right-handed charged-current couplings (A. Stahl in \cite{PichZardoya:1999yv}).

\begin{figure}[h]
\centering
\includegraphics[width=0.5\textwidth]{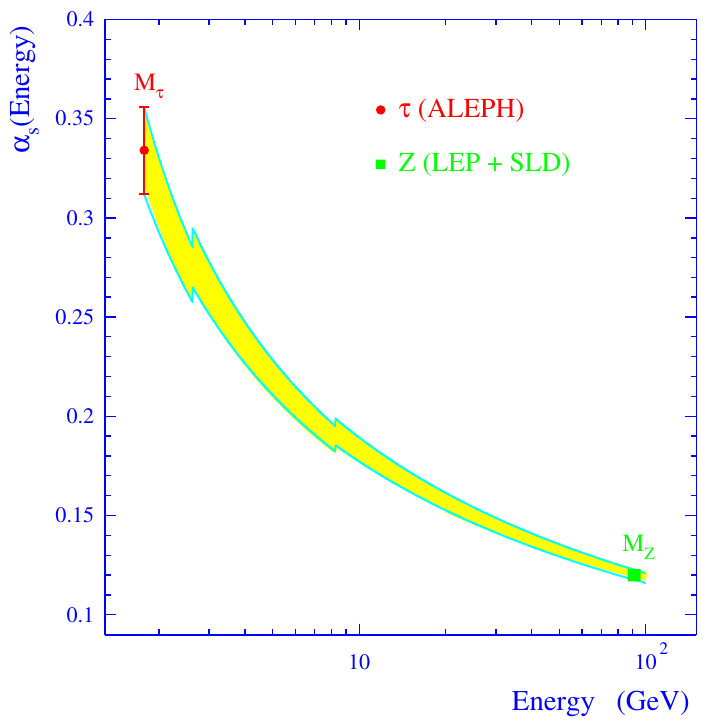}
\caption{Strong coupling extracted from $\tau$ decays, at $\mu=m_\tau$, compared with the $Z$-peak measurement. The band shows the predicted QCD running (M. Davier in \cite{Sobie:2001it}).}
\label{fig:StrongCoupling}
\end{figure}

Detailed experimental analyses of the inclusive hadronic $\tau$ decay width and its invariant-mass distribution, performed by ALEPH (L.~Duflot in \cite{Gan:1992fum}), CLEO and OPAL (S. Menke in \cite{PichZardoya:1999yv}), put on very firm grounds the determination of the strong coupling at the $\tau$ mass (E. Braaten in \cite{Smith:1997iq}),
providing a beautiful test of its QCD running, shown in Fig.~\ref{fig:StrongCoupling}, and a very accurate measurement of $\alpha_s(M_Z)_\tau = 0.1202\pm 0.0027$ (M.~Davier in \cite{Sobie:2001it}), in excellent agreement with the direct determination at the $Z$ peak, $\alpha_s(M_Z)_Z = 0.1183\pm 0.0027$. To put in perspective the importance of these analyses, we should recall that the advocated pre-LEP value was $\alpha_s(M_Z) = 0.11\pm 0.01$ \cite{Altarelli:1990gv}
and the first (1992) precise lattice determination, $\alpha_s(M_Z) = 0.105\pm 0.004$ \cite{El-Khadra:1992ytg}, was substantially lower than the currently accepted value.

\subsection{The 1995 Nobel Prize in Physics}

In 1995, Martin L. Perl (New York 1927 - Palo Alto 2014) was awarded the Nobel Prize in Physics for the discovery of the $\tau$ lepton, sharing the prize with Frederick Reines (Paterson 1918 - Orange 1998) for the first detection of  neutrinos. This high recognition was well celebrated by the whole $\tau$ community. Martin participated very actively in the Tau Physics workshops, being an honorary member of the IAC until the end of his life. He attended in person all workshops from 1990 (Orsay) to 2002 (Santa Cruz), and in 2008 (Novosibirsk),
providing always his very strong support  and wise advice.

\section{B-Factory Era}

The advent of the B factories opened a new era (Victoria 2000 \cite{Sobie:2001it}, Santa Cruz 2002 \cite{Schalk:2002zs}, Nara 2004 \cite{Ohshima:2005iv}, Pisa 2006 \cite{Cei:2007zza}, Novosibirsk 2008 \cite{Bondar:2009zz}, Manchester 2010 \cite{Lafferty:2011zz}), characterised by very large data samples. This allowed for detailed studies of exclusive invariant-mass distributions, resonance structures and high-multiplicity modes (J. Portolés in \cite{Ohshima:2005iv}; H.~Hayashii, M. Fujikawa in \cite{Bondar:2009zz}; A. Adametz, M. Davier, M.J. Lee in \cite{Lafferty:2011zz};  R.J. Sobie in \cite{Hayasaka:2014xya}; H.~Hayashii, E. Tanaka, P.~Roig in \cite{Stahl:2015oci}). 

\begin{figure}[h]
\centering
\includegraphics[width=0.5\textwidth]{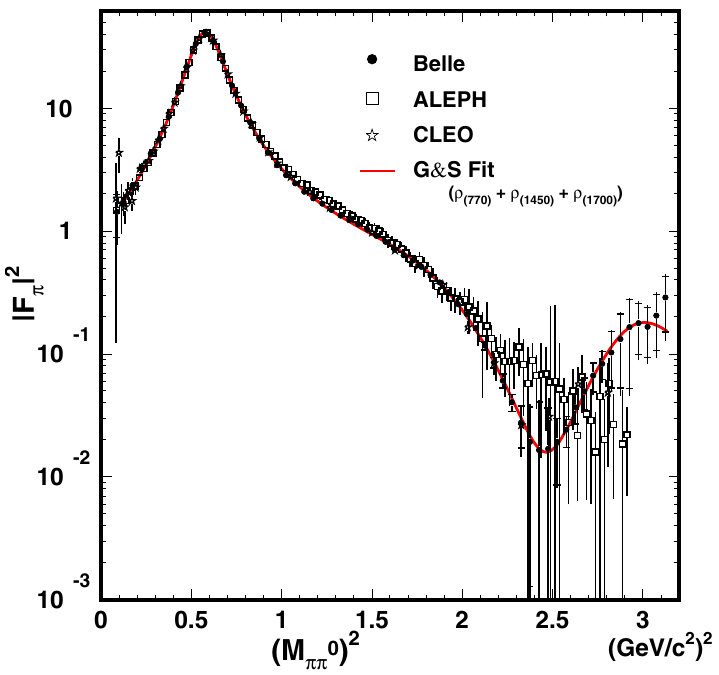}
\caption{Pion form factor in $\tau^-\to\nu_\tau\pi^-\pi^0$ (H. Hayashii, M. Fujikawa in \cite{Bondar:2009zz}).}
\label{fig:PionFF}
\end{figure}

Searches for processes violating lepton flavour or lepton number were highly benefited by the huge available statistics, reaching sensitivities of a few $10^{-8}$ in many $\tau$ decay modes (A.~Cervelli, M. Lewczuk, K. Inami in \cite{Lafferty:2011zz};
Y. Jin, D.A. Epifanov, H. Aihara, S. Eidelman in \cite{Proceedings:2019ihn}).
This has been complemented by the strong constraints coming from $\mu$ decays
(B.~Golden in \cite{Lafferty:2011zz}; H. Natori in \cite{Hayasaka:2014xya})
and $\mu N \to e N$ conversion (T.~Iwamoto in \cite{Proceedings:2019ihn})).

\begin{figure}[h]
\centering
\includegraphics[width=0.9\textwidth]{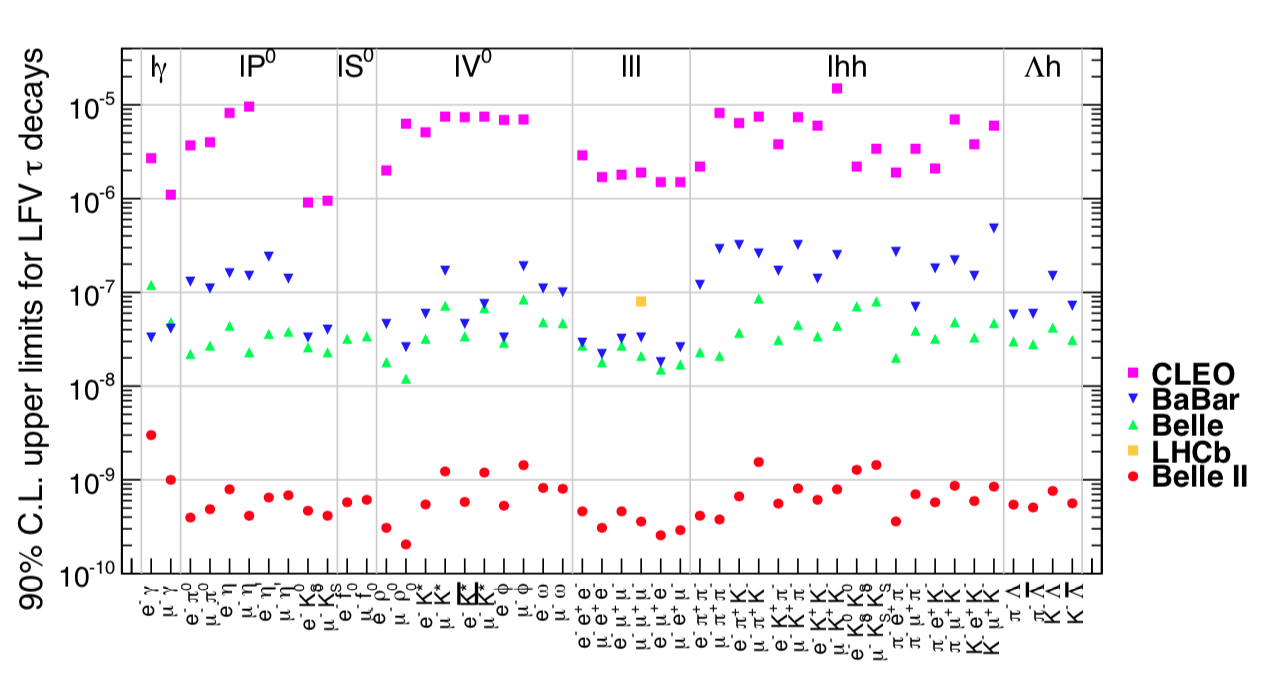}
\caption{Current (90\% C.L.) upper limits on $\tau$ decays violating
lepton flavour or lepton number, and future prospects at Belle-II (M. Hernández Villanueva in \cite{Proceedings:2019ihn}).}
\label{fig:LFV}
\end{figure}

Many analyses of LEP data continued during this period. Worth mentioning are the complete list of ALEPH branching ratios (M. Davier in \cite{Ohshima:2005iv}), and the inclusive strange spectral functions reported by ALEPH (M. Davier et al in \cite{Sobie:2001it}) and OPAL (W. Mader in \cite{Ohshima:2005iv}), which made it possible to extract values of the strange quark mass and the Cabibbo mixing (J. Prades in \cite{Ohshima:2005iv}). I would like to stress here the comprehensive works on SU(3) breaking (and light-by-light contributions to $g-2$) developed by my collaborator and friend Ximo Prades  (Castellón 1963 - Granada 2010), who unfortunately is no longer with us.

A very important theoretical development was the impressive calculation of the $O(\alpha_s^4)$ correction to the inclusive $\tau$ hadronic width (Baikov et al in 
\cite{Bondar:2009zz}), which would be later complemented with a 5-loop computation of the QCD $\beta$ function (Baikov et al in \cite{Yuan:2017tgx}) and an updated version of the Cabibbo-allowed ALEPH spectral functions (Z. Zhang in \cite{Stahl:2015oci}). This has made possible to determine the strong coupling at the N${}^3$LO, triggering a huge theoretical activity and many lively discussions at different $\tau$ meetings (moment analyses, OPE contributions, renormalons, duality violations, etc). The current status has been recently reviewed in \cite{Pich:2020gzz}, which contains an extensive list of relevant references.

The muon anomalous magnetic moment is another timely topic that has been discussed in detail at different meetings. The BNL E821 data was presented by B.L. Roberts in \cite{Schalk:2002zs} and D.~Hertzog in \cite{Ohshima:2005iv}, and the relevant SM contributions have been reviewed: QED (T.~Kinoshita in \cite{Ohshima:2005iv}; M. Hayakawa in \cite{Hayasaka:2014xya}), electroweak (A. Czarnecki in \cite{Schalk:2002zs}), hadronic vacuum polarization (A.~Hoecker in \cite{Schalk:2002zs})
and light-by light (A. Vainshtein in \cite{Cei:2007zza}; E. de Rafael in \cite{Hayasaka:2014xya}; H.~Meyer in \cite{Proceedings:2019ihn}). In particular, there have been many experimental contributions from BaBar, Belle, BES, CLEO, CMD, KEDR, KLOE, SND, etc, providing the necessary input to the dispersive evaluation of the hadronic vacuum polarization contribution  
(M. Davier, H. Hagiwara et al in \cite{Yuan:2017tgx}; B.~Shwartz in \cite{Proceedings:2019ihn}). The radiative return method (J.H. K\"uhn in \cite{Ohshima:2005iv}; G. Rodrigo in \cite{Cei:2007zza})
and the complementary information from $\tau$ decay data were also discussed (S. Eidelman, M.~Davier in \cite{Sobie:2001it}; A. Hoecker in \cite{Schalk:2002zs}; Z. Zhang in \cite{Hayasaka:2014xya}). In the 2021 workshop, we have of course seen the new measurement of the Muon $g-2$ experiment at Fermilab (J. Stapleton) and an overview of the theory status (G.~Colangelo).
The prospects to improve the poorly known electromagnetic dipole moments of the $\tau$ lepton have been also analysed (M. Fael et al in \cite{Hayasaka:2014xya}; M. Hernández-Ruiz et al in \cite{Proceedings:2019ihn}).

The most important achievements in neutrino physics were also presented at the $\tau$ workshops, where a dedicated session has been always scheduled for this. Worth a mention for their direct connection with the $\tau$ lepton are the first direct observation of the $\tau$ neutrino by the DONUT experiment (B. Baller in \cite{Sobie:2001it}), and the first $\nu_\mu\to\nu_\tau$ events registered ten years later with the OPERA detector (Y. Gornushkin in \cite{Lafferty:2011zz}).
The SNO measurement of solar neutrino fluxes (E.W.~Beier in \cite{Schalk:2002zs}) and the SuperKamiokande atmospheric $\nu_\mu\to\nu_\tau$ signal (R. Svoboda in \cite{Sobie:2001it}; J. Shirai in \cite{Ohshima:2005iv}) were, of course, major milestones in neutrino oscillations. Regular updates of the  oscillation data from solar, atmospheric, reactor and accelerator experiments have been discussed since then (J.~Shirai in \cite{Ohshima:2005iv}; C. Howcroft in \cite{Cei:2007zza}; R.A. Johnson in \cite{Hayasaka:2014xya}; G.J.~Barker in \cite{Stahl:2015oci}).

\section{LHC Times}

With the start of operation of the LHC the main focus has moved to the energy frontier (Manchester 2010 \cite{Lafferty:2011zz}, Nagoya 2012 \cite{Hayasaka:2014xya}, Aachen 2014 \cite{Stahl:2015oci}, Beijing 2016 \cite{Yuan:2017tgx}, Amsterdam 2018 \cite{Proceedings:2019ihn}, Indiana 2021). 
The high-momenta $\tau$'s produced at the LHC  turn out to be an excellent signature to probe new physics. They have low multiplicity and good tagging efficiency. Moreover, their decay products are tightly collimated (mini-jet like) and momentum reconstruction is possible. Being a third-generation particle, the $\tau$ is also the lepton that couples more strongly to the Higgs; $H\to\tau^+\tau^-$ is in fact the $4^{\mathrm{th}}$ largest branching ratio of the Higgs boson.

Thus, in the more recent $\tau$ workshops, we have seen a proliferation of Higgs-related measurements and exclusion plots from clever search analyses (D. Chakraborty, S. Knutzen in \cite{Stahl:2015oci}; A.~Lusiani, Z. Mao, R. Reece in \cite{Yuan:2017tgx}; C. Caputo, F. Lyu in \cite{Proceedings:2019ihn}). The detection of $H\to\tau^+\tau^-$ events and the corresponding measurement of the $\tau$ Yukawa coupling (T. Müller in \cite{Stahl:2015oci}; D.~Zanzi in \cite{Yuan:2017tgx}; L.~Schildgen in \cite{Proceedings:2019ihn}) have been major milestones, together with the limits set on lepton-flavour-violating couplings of the Higgs (A. Nehrkorn in \cite{Yuan:2017tgx}; B. Le in \cite{Proceedings:2019ihn}) and the $Z$ boson (K. De Bruyn, A. Nehrkorn in \cite{Yuan:2017tgx}; W.S. Chan in \cite{Proceedings:2019ihn}). It is remarkable that the LHC bounds on $Z\to\ell\ell'$ with $\ell\not=\ell'$ are already better than the LEP ones. The LHC experiments have also provided relevant bounds on the
$\tau\to 3\mu$ decay mode (K. De Bruyn in \cite{Yuan:2017tgx}).

\begin{figure}[h]
\centering
\includegraphics[width=0.72\textwidth]{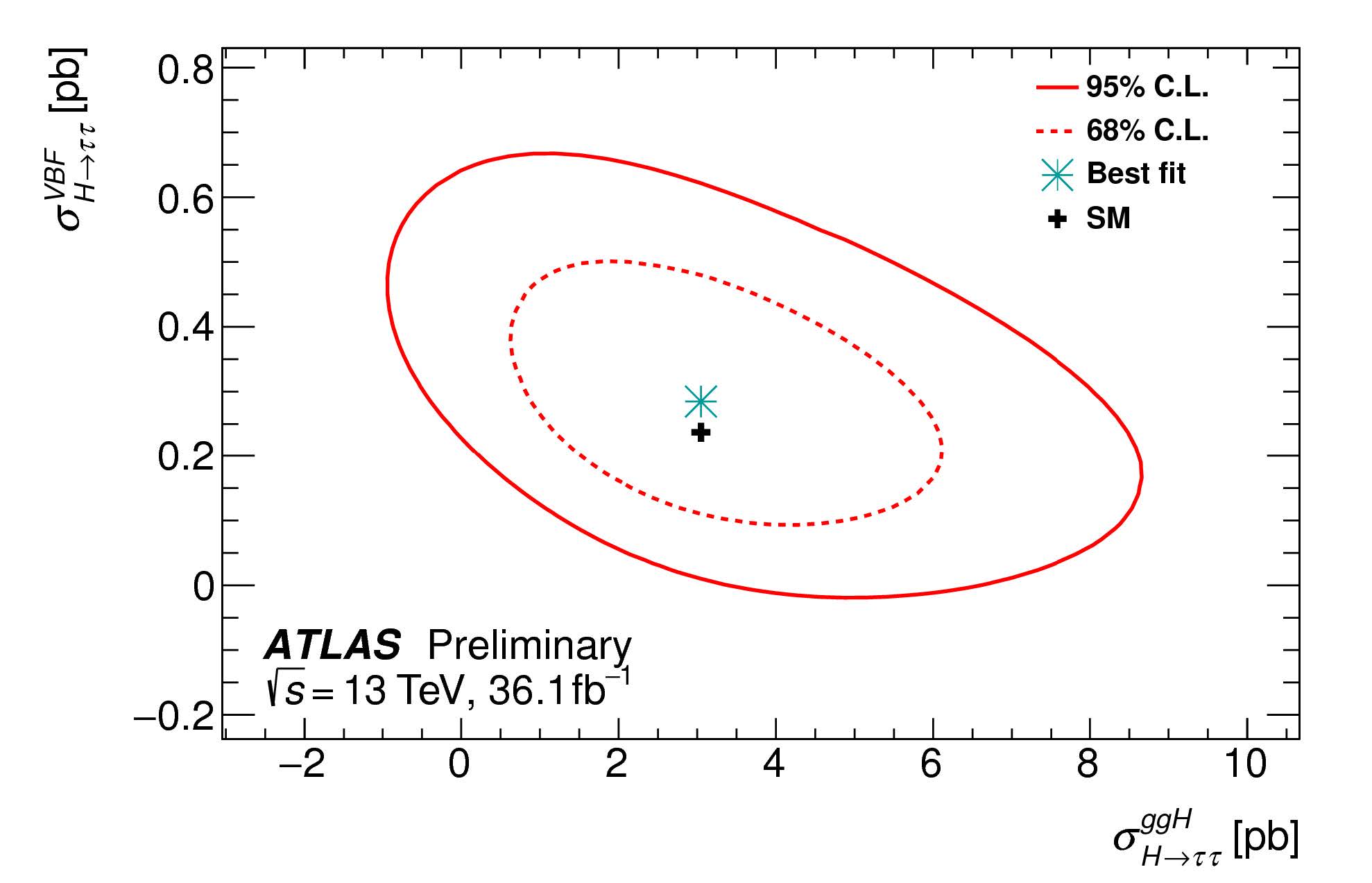}
\caption{ATLAS measurement of $\sigma(H\to\tau^+\tau^-)$ (L.~Schildgen in \cite{Proceedings:2019ihn}).}
\label{fig:H2tau}
\end{figure}

The strong improvement achieved in $\tau$ detection techniques has made possible to perform relevant tests of the SM itself. Worth mentioning are the first hadron-collider measurement of the $\tau$ polarization in $W\to\tau\nu$ decays (Z. Czyczula in \cite{Hayasaka:2014xya}) and the 
$\tau$ polarization asymmetry in $Z\to\tau^+\tau^-$ (V. Cherepanov in \cite{Yuan:2017tgx}). More recently, the ATLAS and CMS experiments have been able to test lepton universality in $W\to\ell\nu_\ell$ decays, in good agreement with the SM, clarifying the puzzling $2.5\,\sigma$ excess of $\tau$ events observed a long time ago in the LEP data.

The flavour anomalies identified in $B$ decays have been one of the more recent highlights (E.~Manoni in \cite{Hayasaka:2014xya}; A. Celis, T. Kuhr in \cite{Stahl:2015oci}; K. De Bruyn, S. Hirose, X.-Q. Li in \cite{Yuan:2017tgx}; S.~Benson, S. Fajfer in \cite{Proceedings:2019ihn}), since they indicate unexpected large violations of lepton universality in $b\to c\tau\nu$ and $b\to s\mu^+\mu^-$. The available high-$p_T$ data on di-tau production at the LHC provides complementary information, constraining many suggested new-physics scenarios (D.A.~Faroughy in \cite{Proceedings:2019ihn}).

\begin{figure}[h]
\centering
\includegraphics[width=0.7\textwidth]{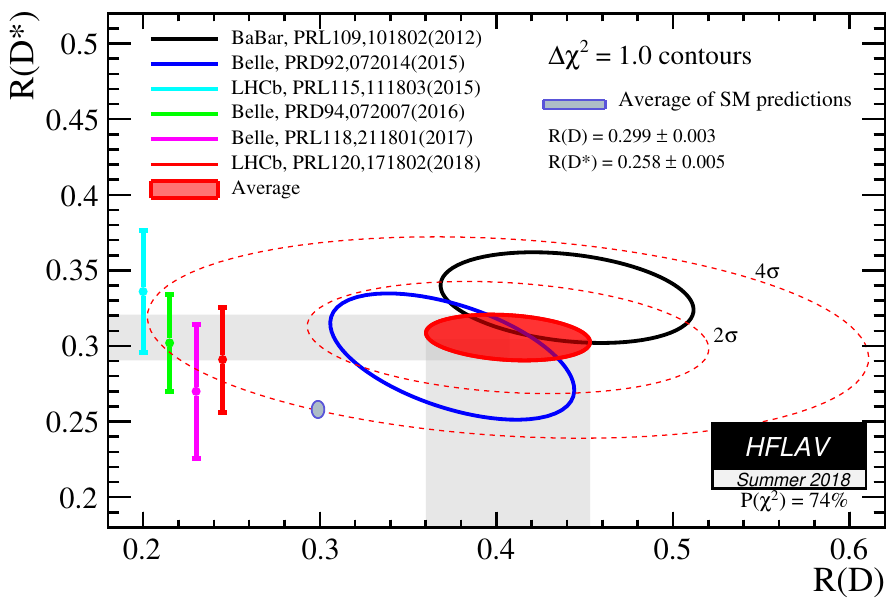}
\caption{Measurements of the ratios $R_{D^{(*)}} = \Gamma[B\to D^{(*)}\tau\nu]/\Gamma[B\to D^{(*)}\ell\nu]$ ($\ell = e,\mu$),
compared with the SM predictions \cite{HFLAV:2019otj} (S.~Benson in \cite{Proceedings:2019ihn}).}
\label{fig:RD}
\end{figure}

Another surprising result was reported by the BaBar collaboration (R. Sobie in \cite{Hayasaka:2014xya}), who observed a CP-violating rate asymmetry in the decay $\tau^-\to\nu_\tau\pi^-K_S^0\, (\ge 0 \pi^0)$ that deviates by $2.8\,\sigma$ from the SM expectation for $K^0-\bar{K}^{0}$ mixing. The BaBar signal has not been confirmed by Belle, which did not reach the required sensitivity (M. Hernández Villanueva in \cite{Proceedings:2019ihn}), and seems incompatible with other sets of flavour data (V. Cirigliano et al 
in \cite{Proceedings:2019ihn}).
While future measurements should clarify this situation, the search for signatures of CP violation in $\tau$ decays remains an interesting goal (I. Bigi in \cite{Proceedings:2019ihn}).

The most important achievements in Astroparticle physics have been also reviewed at the $\tau$ workshops. Two recent IceCube highlights are the discovery of an astrophysical neutrino flux in 2013, which marked the birth of neutrino astronomy (D. Xu in \cite{Yuan:2017tgx}), and the evidence for the identification of a blazar as an astrophysical neutrino source reported in 2018 (D. van Eijk in \cite{Proceedings:2019ihn}).

\section{Future Prospects}

The forthcoming high-statistics data samples that will soon be accumulated by the Belle-II detector will give a new boost to precision $\tau$ physics \cite{Belle-II:2018jsg}. In addition to much larger sensitivities to decays violating the lepton flavour or the lepton number, one expects significant improvements on the $\tau$ lifetime and branching ratios, decay distributions, CP asymmetries, Michel parameters, etc (M.~Hernández Villanueva in \cite{Proceedings:2019ihn}). This superb physics potential will be complemented with more precise measurements of the $\tau$ mass at BES-III (J. Zhang in  \cite{Proceedings:2019ihn}), and a new generation of muon experiments (C. Wu in \cite{Yuan:2017tgx}; R. Bonventre, A. Bravar, A. Driutti, T.~Iwamoto, A.-K. Perrevoort, N. Teshima in \cite{Proceedings:2019ihn}), neutrinoless double-beta-decay searches (L. Cardani in \cite{Proceedings:2019ihn})
and neutrino detectors (M. Komatsu, Z. Li, H. Lu in \cite{Yuan:2017tgx}; D.~van Eijk, I.~Esteban, P.~Fernández, A. Pocar et al, H.~Seo, C. Timmermans, A. Tonazzo, M. Trini in \cite{Proceedings:2019ihn}) at different laboratories.

The LHC is also going to start its new Run 3, aiming to a sizeable increment of the integrated luminosity. This will be followed later by a much more significant  improvement of the instantaneous luminosity at the HL-LHC, which will increase the potential for new discoveries \cite{Cerri:2018ypt}. In the long term, several linear and circular high-energy colliders are being discussed. Huge and clean $\tau^+\tau^-$ data samples could be provided by an electron-positron TeraZ facility, running at the $Z$ peak (M. Dam in \cite{Proceedings:2019ihn}). The projects to build a high-luminosity super-$\tau$cF  (S. Eidelman in \cite{Stahl:2015oci}) are also in a quite advanced stage.

Thus, there is a bright future ahead of us with lots of interesting physics to be explored. We can look forward to many relevant experimental discoveries to be celebrated at the $40^{\mathrm{th}}$ Tau Lepton Physics anniversary in 2030.

\section*{Acknowledgements}
I would like to thank Michel Davier for his continuous support to the $\tau$ workshops, 
and the local organizers for making possible this $16^{\mathrm{th}}$ meeting, in spite of the difficult circumstances. 
This work has been supported by 
MCIN/AEI/10.13039/501100011033, Grant No. PID2020-114473GB-I00,
and by the Generalitat Valenciana, Grant No. Prometeo/2021/071.


%

\bibliography{MyBibFile}

\begin{thebibliography}{10}
\providecommand{\url}[1]{\texttt{#1}}
\providecommand{\urlprefix}{URL }
\expandafter\ifx\csname urlstyle\endcsname\relax
  \providecommand{\doi}[1]{doi:\discretionary{}{}{}#1}\else
  \providecommand{\doi}{doi:\discretionary{}{}{}\begingroup
  \urlstyle{rm}\Url}\fi
\providecommand{\eprint}[2][]{\url{#2}}

\bibitem{Davier:1991jq}
M.~Davier and B.~Jean-Marie, eds.,
\newblock \emph{{Proceedings, Tau Lepton Physics (TAU 90)}: {Orsay, France,
  September 24-27, 1990}}. Editions Frontieres, Gif-sur-Yvette Cedex, France
  (1991).

\bibitem{Gan:1992fum}
K.~K. Gan, ed.,
\newblock \emph{{Proceedings, 2nd Workshop on Tau Lepton Physics (TAU 92)}:
  {Columbus, Ohio, USA, September 8-11, 1992}}. World Scientific, New Jersey
  (1992).

\bibitem{Rolandi:1995yx}
G.~Rolandi, ed.,
\newblock \emph{{Proceedings, 3rd Workshop on Tau Lepton Physics (TAU 94):
  Montreux, Switzerland, September 19-22, 1994}}. Nucl. Phys. B (Proc. Suppl.)
  40 (1995).

\bibitem{Smith:1997iq}
J.~G. Smith and W.~Toki, eds.,
\newblock \emph{{Proceedings, 4th Workshop on Tau Lepton Physics (TAU 96):
  Estes Park, Colorado, USA, September 16-19, 1996}}. Nucl. Phys. B (Proc.
  Suppl.) 55C (1997).

\bibitem{PichZardoya:1999yv}
A.~Pich~Zardoya and A.~Ruiz~Jimeno, eds.,
\newblock \emph{{Proceedings, 5th Workshop on Tau Lepton Physics (TAU 98):
  Santander, Spain, September 14-17, 1998}}. Nucl. Phys. B (Proc. Suppl.) 76
  (1999).

\bibitem{Sobie:2001it}
R.~J. Sobie and J.~M. Roney, eds.,
\newblock \emph{{Proceedings, 6th International Workshop on Tau Lepton Physics
  (TAU 00): Victoria, Canada, September 18-21, 2000}}. Nucl. Phys. B (Proc.
  Suppl.) 98 (2001).

\bibitem{Schalk:2002zs}
T.~Schalk and A.~Seiden, eds.,
\newblock \emph{{Proceedings, 7th International Workshop on Tau Lepton Physics
  (TAU 02): Santa Cruz, California, USA, September 10-13, 2002}}. Nucl. Phys. B
  (Proc. Suppl.) 123 (2002).

\bibitem{Ohshima:2005iv}
T.~Ohshima and H.~Hayashii, eds.,
\newblock \emph{{Proceedings, 8th International Workshop on Tau Lepton Physics
  (TAU 04): Nara, Japan, September 14-17, 2004}}. Nucl. Phys. B (Proc. Suppl.)
  144 (2005).

\bibitem{Cei:2007zza}
F.~Cei, I.~Ferrante and A.~Lusiani, eds.,
\newblock \emph{{Proceedings, 9th International Workshop on Tau Lepton Physics
  (TAU 06): Pisa, Italy, September 19-22, 2006}}. Nucl. Phys. B (Proc. Suppl.)
  169,
\newblock \doi{10.1016/S0920-5632(07)00392-1} (2007).

\bibitem{Bondar:2009zz}
A.~Bondar and S.~Eidelman, eds.,
\newblock \emph{{Proceedings, 10th International Workshop on Tau Lepton Physics
  (TAU 09): Novosibirsk, Russia, 22-25 September 2008}}. Nucl. Phys. B (Proc.
  Suppl.) 189 (2009).

\bibitem{Lafferty:2011zz}
G.~Lafferty and S.~Soldner-Rembold, eds.,
\newblock \emph{{Proceedings, 11th International Workshop on Tau Lepton Physics
  (TAU 10): Manchester, UK, September 13-17, 2010}}. Nucl. Phys. B (Proc.
  Suppl.) 218 (2011).

\bibitem{Hayasaka:2014xya}
K.~Hayasaka and T.~Iijima, eds.,
\newblock \emph{{Proceedings, 12th International Workshop on Tau Lepton Physics
  (TAU 12): Nagoya, Japan, September 17-21, 2012}}. Nucl. Phys. B (Proc.
  Suppl.) 253-255 (2014).

\bibitem{Stahl:2015oci}
A.~Stahl and I.~M. Nugent, eds.,
\newblock \emph{{Proceedings, 13th International Workshop on Tau Lepton Physics
  (TAU 14): Aachen, Germany, September 15-19, 2014}}. Nucl. Phys. B (Proc.
  Suppl.) 260 (2015).

\bibitem{Yuan:2017tgx}
C.~Yuan, X.~Mo and L.~Wang, eds.,
\newblock \emph{{Proceedings, 14th International Workshop on Tau Lepton Physics
  (TAU 16): Beijing, China, September 19-23, 2016}}. Nucl. Phys. B (Proc.
  Suppl.) 287-288 (2017).

\bibitem{Proceedings:2019ihn}
O.~Igonkina and M.~Morgenstern, eds.,
\newblock \emph{{Proceedings, 15th International Workshop on Tau Lepton Physics
  (TAU 18): Amsterdam, Netherlands, September 24-28, 2018}}. SciPost Phys.
  Proc. 1 (2019).

\bibitem{Pich:2013lsa}
A.~Pich,
\newblock \emph{{Precision Tau Physics}},
\newblock Prog. Part. Nucl. Phys. \textbf{75}, 41 (2014),
\newblock \doi{10.1016/j.ppnp.2013.11.002},
\newblock \eprint{1310.7922}.

\bibitem{Perl:1975bf}
M.~L. Perl \emph{et~al.},
\newblock \emph{{Evidence for Anomalous Lepton Production in $e^+ - e^-$
  Annihilation}},
\newblock Phys. Rev. Lett. \textbf{35}, 1489 (1975),
\newblock \doi{10.1103/PhysRevLett.35.1489}.

\bibitem{Tsai:1971vv}
Y.-S. Tsai,
\newblock \emph{{Decay Correlations of Heavy Leptons in $e^+ + e^- \to\ell^+
  +\ell^-$}},
\newblock Phys. Rev. D \textbf{4}, 2821 (1971),
\newblock \doi{10.1103/PhysRevD.13.771},
\newblock [Erratum: Phys.Rev.D 13, 771 (1976)].

\bibitem{Barish:1987nj}
B.~C. Barish and R.~Stroynowski,
\newblock \emph{{The Physics of the tau Lepton}},
\newblock Phys. Rept. \textbf{157}, 1 (1988),
\newblock \doi{10.1016/0370-1573(88)90103-2}.

\bibitem{Derrick:1987sp}
M.~Derrick \emph{et~al.},
\newblock \emph{{Evidence for the Decay $\tau^+ \to \pi^+ \eta
  \bar{\nu}_\tau$}},
\newblock Phys. Lett. B \textbf{189}, 260 (1987),
\newblock \doi{10.1016/0370-2693(87)91308-6}.

\bibitem{Proceedings:1989lya}
L.~V. Beers, ed.,
\newblock \emph{{Proceedings, Tau - Charm Factory Workshop: Study of Tau, Charm
  and $J/\psi$ Physics, Development of High Luminosity $e^+ e^-$ Rings, Design
  of $e^+ e^-$ Detectors for Tau Charm Physics. SLAC-Report-343}} (1989).

\bibitem{Narison:1988ni}
S.~Narison and A.~Pich,
\newblock \emph{{QCD Formulation of the tau Decay and Determination of
  $\Lambda_{\overline{\mathrm{MS}}}$}},
\newblock Phys. Lett. B \textbf{211}, 183 (1988),
\newblock \doi{10.1016/0370-2693(88)90830-1}.

\bibitem{ParticleDataGroup:2020ssz}
P.~A. Zyla \emph{et~al.},
\newblock \emph{{Review of Particle Physics}},
\newblock PTEP \textbf{2020}(8), 083C01 (2020),
\newblock \doi{10.1093/ptep/ptaa104}.

\bibitem{Altarelli:1990gv}
G.~Altarelli,
\newblock \emph{{QCD and experiment}},
\newblock In \emph{{15th APS Division of Particles and Fields General
  Meeting}}, pp. 170--206 (1990).

\bibitem{El-Khadra:1992ytg}
A.~X. El-Khadra, G.~Hockney, A.~S. Kronfeld and P.~B. Mackenzie,
\newblock \emph{{A Determination of the strong coupling constant from the
  charmonium spectrum}},
\newblock Phys. Rev. Lett. \textbf{69}, 729 (1992),
\newblock \doi{10.1103/PhysRevLett.69.729}.

\bibitem{Pich:2020gzz}
A.~Pich,
\newblock \emph{{Precision physics with inclusive QCD processes}},
\newblock Prog. Part. Nucl. Phys. \textbf{117}, 103846 (2021),
\newblock \doi{10.1016/j.ppnp.2020.103846},
\newblock \eprint{2012.04716}.

\bibitem{HFLAV:2019otj}
Y.~S. Amhis \emph{et~al.},
\newblock \emph{{Averages of b-hadron, c-hadron, and $\tau $-lepton properties
  as of 2018}},
\newblock Eur. Phys. J. C \textbf{81}(3), 226 (2021),
\newblock \doi{10.1140/epjc/s10052-020-8156-7},
\newblock \eprint{1909.12524}.

\bibitem{Belle-II:2018jsg}
W.~Altmannshofer \emph{et~al.},
\newblock \emph{{The Belle II Physics Book}},
\newblock PTEP \textbf{2019}(12), 123C01 (2019),
\newblock \doi{10.1093/ptep/ptz106},
\newblock [Erratum: PTEP 2020, 029201 (2020)],
\newblock \eprint{1808.10567}.

\bibitem{Cerri:2018ypt}
A.~Cerri \emph{et~al.},
\newblock \emph{{Report from Working Group 4}: {Opportunities in Flavour
  Physics at the HL-LHC and HE-LHC}},
\newblock CERN Yellow Rep. Monogr. \textbf{7}, 867 (2019),
\newblock \doi{10.23731/CYRM-2019-007.867},
\newblock \eprint{1812.07638}.

\end{thebibliography}

\nolinenumbers

\end{document}